\documentclass[12pt,preprint]{aastex}
\usepackage{natbib}
\usepackage{graphicx}
\shorttitle{Severe challenge to MOND phenomenology}
\begin{document}
\title{A severe challenge to the MOND phenomenology in our Galaxy}
\author{Man Ho Chan \& Ka Chung Law}
\affil{Department of Science and Environmental Studies, The Education University of Hong Kong, Hong Kong, China}
\email{chanmh@eduhk.hk}

\begin{abstract}
Modified Newtonian Dynamics (MOND) is one of the most popular alternative theories of dark matter to explain the missing mass problem in galaxies. Although it remains controversial regarding MOND as a fundamental theory, MOND phenomenology has been shown to widely apply in different galaxies, which gives challenges to the standard $\Lambda$ cold dark matter model. In this article, we derive analytically the galactic rotation curve gradient in the MOND framework and present a rigorous analysis to examine the MOND phenomenology in our Galaxy. By assuming a benchmark baryonic disk density profile and two popular families of MOND interpolating functions, we show for the first time that the recent discovery of the declining Galactic rotation curve in the outer region ($R \approx 17-23$ kpc) can almost rule out the MOND phenomenology at more than $5\sigma$. This strongly supports some of the previous studies claiming that MOND is neither a fundamental theory nor a universal description of galactic properties.
\end{abstract}

\keywords{Modified Gravity, Galaxy}

\section{Introduction}
Dark matter theory and modified gravity theory are two kinds of competing theories to explain the missing mass problem in galaxies and galaxy clusters \citep{Bertone}. The controversy between these two theories is still on-going \citep{Martens}. The successful predictions based on the standard $\Lambda$ cold dark matter ($\Lambda$CDM) model in large-scale structures have gained important credence on dark matter theory \citep{Croft,Spergel,Peacock} while the galactic relations (e.g. baryonic Tully-Fisher relation, radial acceleration relation) show a certain support to modified gravity theory \citep{McGaugh,Islam}. 

As one of the earliest versions of modified gravity theory, Modified Newtonian Dynamics (MOND) has been examined for four decades. It is well-known that MOND works very well in galaxies but it gives poor fits with the data of galaxy clusters \citep{Sanders,Chan4}. Although it is dubious for MOND to be a fundamental theory \citep{Chan}, the phenomenological description using MOND agrees with the data in galaxies. For example, the baryonic Tully-Fisher relation \citep{Lelli} and the radial acceleration relation \citep{McGaugh2} in galaxies give excellent agreement with the MOND's predictions \citep{McGaugh,Islam}. Also, the apparent flat rotation curves (RCs) in most of the galaxies match the MOND behavior at large distance from the galactic centres \citep{Sanders,Gentile}. Therefore, these potentially suggest that MOND can be regarded as a universal phenomenological description in galaxies (the MOND phenomenology) \citep{Gentile,Dutton}. Many studies have used the MOND phenomenology to study galactic dark matter and galactic scaling relations \citep{Gentile2,Ho}. However, most of the galactic data have large uncertainties. For example, the high-quality rotation curve data (SPARC) used in supporting MOND have an average of 7.1\% observational uncertainty (not including systematic uncertainties) \citep{Lelli2}. Therefore, it is very difficult for us to rigorously verify or falsify MOND as a universal phenomenological description on galactic scale.

On the other hand, recent high-quality observations of the Milky Way rotation curve (MWRC) data in the outer region ($R>17$ kpc) give very small marginal errors (with an average of 2-3\% observational uncertainty only) in the rotation curve fittings \citep{Eilers,Wang,Labini,Ou}. The MWRC shows a very clear decreasing trend from $V \approx 220$ km/s to $V \approx 170$ km/s in the range of $R=12-27$ kpc \citep{Labini,Ou}. The decreasing RC may even extend to 200 kpc based on the data of the Milky Way satellites, although the decreasing rate is relatively smaller than that in the range of $R=12-27$ kpc \citep{Vasiliev,Wang}. Since MOND generally suggests a flat rotation curve for large $R$, such a decline in MWRC is not trivial for MOND. However, statistical margins might still allow an almost flat rotation curve to fit the decline in MWRC due to the measurement uncertainties. 

In view of this problem, a more precise way is to focus particularly on the RC gradient $dV/dR$ to investigate MOND. Earlier measurements give $dV/dR=-(1.7 \pm 0.1)$ km/s/kpc at large $R$ \citep{Eilers}, which can still be explained by the MOND behavior \citep{McGaugh3,McGaugh6}. However, more recent measurements and analyses give $dV/dR=-(2.3 \pm 0.2)$ km/s/kpc \citep{Wang2} and even $dV/dR<-3$ km/s/kpc \citep{Labini,Ou} at large $R$, which might give challenges to the MOND phenomenological description. 

In this article, we present a rigorous analysis to understand how large of $dV/dR$ can MOND tolerate in galaxies. We specifically use the MWRC data to examine the MOND phenomenology using two major families of MOND interpolating functions.  By assuming a benchmark baryonic disk density profile, we show that the MWRC data can rule out MOND phenomenology for the two major families of interpolating functions at more than $5\sigma$, which can almost exhaustively represent all possible variations of MOND. This suggests that there is likely no room for MOND phenomenology in our Galaxy, which provides the first severe challenge to the alleged universal nature of MOND phenomenology on galactic scale.

\section{The rotation curve behavior in the MOND framework}
MOND suggests a modification of gravity if the test particle is moving with an acceleration smaller than a critical value $a_0$. When the normal Newtonian gravitational acceleration $g_N \ll a_0$ (i.e. the deep-MOND regime), the gravity would be modified to \citep{Sanders}
\begin{equation}
g=\sqrt{g_Na_0}.
\end{equation}
When $g_N \gg a_0$ (i.e. the Newtonian regime), the gravity would restore back to the Newtonian description: $g=g_N$. One can imagine that there is a transition from $g=\sqrt{g_Na_0}$ to $g=g_N$ when $g_N$ is changing from a very small value to a large value. In the MOND framework, the transition is described by the interpolating function $\nu$ \citep{Famaey}. Therefore, the modification of gravity can be generally re-written as 
\begin{equation}
g=\nu(g_N/a_0)g_N,
\end{equation}
where $\nu(z)=1$ when $z \rightarrow \infty$ and $\nu(z)=\sqrt{1/z}$ when $z \rightarrow 0$. Although there is no theoretical prediction of $\nu$ based on MOND, there are some suggested families of functions which can satisfy the required transition \citep{Famaey}:
\begin{equation}
\nu(z)=\left[ \frac{1+(1+4z^{-p})^{1/2}}{2} \right]^{1/p},
\end{equation}
\begin{equation}
\nu(z)=[1-\exp(-z^{\delta/2})]^{-1/\delta}.
\end{equation}
Here, the parameters $p$ and $\delta$ are positive real numbers. For the $p$-family, $p=1$ and $p=2$ are commonly known as the simple interpolating function and the standard interpolating function respectively \citep{Famaey,Dutton}. For the $\delta$-family, $\delta=1$ is the well-known function which provides the best description of the radial acceleration relation in galaxies \citep{McGaugh2,Dutton}. In any case, different possible values of $p$ and $\delta$ can almost exhaustively represent all possible ways of transition between the deep-MOND and Newtonian regimes \citep{Dutton}. Although most of the previous studies took $p \ge 1$ and $\delta \ge 1$ \citep{McGaugh5,Dutton}, we generally allow that $p$ and $\delta$ can be any positive real numbers first.

If dark matter does not exist, the gravity is fully contributed by the baryonic matter. In a typical galaxy, the baryonic matter is dominated by the bulge component and the disk component. Since we are concerning about the outer region of a galaxy (i.e. large $R$), the bulge component can be well-approximated by a point-mass function at the Galactic centre with the bulge mass $M_b$. For the disk component, early observations have shown that the disk surface brightness of many galaxies can be well-approximated by an exponential function, including our Galaxy \citep{Freeman,Juric,deJong,McGaugh4}. Therefore, most of the past studies and simulations have used an exponential function to model the surface mass density of the disk component in our Galaxy \citep{Misiriotis,Sofue,Bovy,Licquia}:
\begin{equation}
\Sigma=\Sigma_0e^{-R/a_d},
\end{equation}
where $a_d$ is the disk scale length and $\Sigma_0$ is the central surface mass density. The total disk mass can be given by $M_d=2\pi a_d^2\Sigma_0$. The Newtonian baryonic RC (without MOND transformation) is given by \citep{Freeman,Sofue}
\begin{equation}
V_{\rm bar}(x)=\sqrt{Ax^2\left[I_0(x)K_0(x)-I_1(x)K_1(x) \right]+\frac{B}{x}},
\end{equation}
where $A=2GM_d/a_d$, $B=GM_b/(2a_d)$, $x=R/(2a_d)$, $I_n$ and $K_n$ are the modified Bessel functions of the $n^{\rm th}$ kind.

To make the analysis more comprehensive, apart from following the benchmark exponential function to model the disk component, we will consider an extreme case in which all of the baryonic disk mass is concentrated in $R<17$ kpc (i.e. the concentrated baryonic disk model). This would represent a Keplerian decrease in baryonic RC contribution and hence contribute to a more negative RC slope. For this extreme scenario, the Newtonian baryonic RC without MOND transformation is
\begin{equation}
V_{\rm bar}(x)=\sqrt{\frac{A}{8x}+\frac{B}{x}}.
\end{equation}

Under the framework of MOND, we can transform the Newtonian baryonic RC to the apparent RC $V(R)$ (i.e. the actual rotation curve observed) by using Eq.~(2) as
\begin{equation}
V(x)=\sqrt{\nu \left(\frac{V_{\rm bar}^2(x)}{2a_0a_dx} \right)}V_{\rm bar}(x).
\end{equation}
To understand the behavior of the RC, we take the derivative on $V(x)$ to get the analytic forms of the RC gradient. For the $p$-family, by substituting Eq.~(3) into Eq.~(8), we can get
\begin{equation}
V(x)=2^{-1/2p}V_{\rm bar} \left(1+\sqrt{1+4y^p} \right)^{1/2p},
\end{equation}
where $y=2a_0a_dx/V_{\rm bar}^2$. Taking the derivative on both sides, we can get the RC gradient for the $p$-family:
\begin{equation}
\frac{dV}{dx}=\frac{(\sqrt{1+4y^p}+1+2y^p)V'_{\rm bar}+2y^{p-1}a_0a_dV_{\rm bar}^{-1}}{\sqrt{2^{1/p}(1+\sqrt{1+4y^p})^{2-1/p}(1+4y^p)}}.
\end{equation}
For the $\delta$-family, we can write the apparent RC explicitly:
\begin{equation}
V(x)=\sqrt{\left[1-\exp \left(-y^{-\delta/2} \right) \right]^{-1/\delta}}V_{\rm bar}.
\end{equation}
Similarly, by taking the derivative on both sides, we get
\begin{equation}
\frac{dV}{dx}=\left[1-\exp \left(-y^{-\delta/2} \right) \right]^{-1/2\delta} \left[V'_{\rm bar}+\frac{V_{\rm bar}^2y^{1-\delta/2}}{4\exp(y^{-\delta/2})-4} \times \frac{V_{\rm bar}-2xV'_{\rm bar}}{2a_0a_dx^2} \right].
\end{equation}
Here, $V'_{\rm bar}$ is the derivative on the Newtonian baryonic RC. For the benchmark and the concentrated baryonic disk models, we have
\begin{eqnarray}
V'_{\rm bar}&=&\frac{1}{2V_{\rm bar}(x)} \left[2Ax \left( I_0(x)K_0(x)+xI_1(x)K_0(x) \right. \right. \nonumber\\
&& \left. \left. -xI_0(x)K_1(x) \right)-\frac{B}{x^2} \right],
\end{eqnarray}
and
\begin{equation}
V'_{\rm bar}=-\frac{1}{2V_{\rm bar}} \left(\frac{A}{8x^2}+\frac{B}{x^2} \right)
\end{equation}
respectively. After grouping the appropriate terms in Eq.~(12), we get the RC gradient for the $\delta$-family:
\begin{equation}
\frac{dV}{dx}=\frac{[\exp(y^{-\delta/2})-1-0.5y^{-\delta/2}]V'_{\rm bar}+0.25y^{-\delta/2}V_{\rm bar}x^{-1}}{ [1-\exp(-y^{-\delta/2})]^{1/2\delta}[\exp(y^{-\delta/2})-1]}.
\end{equation}

\section{Data analysis}
We test the MOND RC behavior by using the MWRC data. We focus on the outer region $R \sim 20$ kpc of our Galaxy because MOND effect is significant when $R$ is large. Recent accurate measurements of the MWRC at $R=17-27$ kpc can give a rigorous examination of MOND. The most recent robust analysis of the MWRC combining the data of APOGEE DR17, GAIA, 2MASS and WISE gives a large decline in the RC gradient \citep{Ou}. However, since the systematic uncertainties of the data are very large ($\sim 15$\%) for $R>22.5$ kpc \citep{Ou}, we will only analyze the data within the region $R=17.21-22.27$ kpc (systematic uncertainties $\sim 1-5$\% only). Simple regression of the RC data points gives $dV/dR=(-4.567 \pm 0.532)$ km/s/kpc for $R=17.21-22.27$ kpc. If we include the consideration of the small observational uncertainties for each RC data point, the best-fit RC slope with $5\sigma$ margins is $dV/dR=-5.07^{+2.51}_{-2.49}$ km/s/kpc (99.99994\% C.L.) for $R=17.21-22.27$ kpc (see Fig.~1). The $5\sigma$ margins are calculated using the $\chi^2$ method. By examining different values of $dV/dR$ in the fits, the values of the $5\sigma$ margins can be determined when the $\chi^2$ value is larger than the $5\sigma$ critical value $\chi_{\rm crit}^2=48$ (degrees of freedom = 10). Note that we did not include the systematic uncertainties of the RC data points in determining the RC gradient. The actual effects of the systematic uncertainties on the RC gradient depend on the involved factors. We will discuss these effects in the discussion section.

From Eq.~(10) and Eq.~(15), we can see that the RC gradient depends on the baryonic parameters ($M_b$, $M_d$ and $a_d$) and the MOND parameters ($a_0$, $p$ and $\delta$). In the followings, we take the standard value of $a_0=1.2 \times 10^{-8}$ cm/s$^2$ to perform our analysis \citep{McGaugh2}. For the baryonic parameters, we first assume the fiducial values adopted in \citet{Ou}: $M_b=1.55 \times 10^{10}M_{\odot}$, $M_d=3.65 \times 10^{10}M_{\odot}$ and $a_d=2.35$ kpc. These values are determined by the direct baryonic observations \citep{Misiriotis}, which are somewhat representative and good for performing the preliminary analysis. We will also test a reasonable range of each parameter in the next step.

In Fig.~2, we plot $dV/dR$ as a function of $p$ and $\delta$ for the $p$-family and $\delta$-family interpolating functions respectively at $R=19.71$ kpc. We have taken the data point at $R=19.71$ as the central reference position because it is the median $R$ and the mean value of $R$ among the data points within the range $R=17.21-22.27$ kpc. Generally speaking, our results do not depend on the choice of the central reference position. We can see that larger values of $dV/dR$ (more positive or less negative) would result when $p$ and $\delta$ are larger. For the benchmark baryonic disk model, the $5\sigma$ allowed ranges of $p$ and $\delta$ are $p=0.19-0.46$ and $\delta=0.17-0.43$ respectively. However, the Mercury precession data have constrained $p \ge 1.22$ and $\delta \ge 0.33$ \citep{Chan2}. Therefore, a large range of $\delta$ and all values of $p$ are ruled out based on the Mercury precession data and MWRC data. In particular, the cases with $p=1$, $p=2$ and $\delta=1$ can give good agreements with the data of galactic rotation curves \citep{McGaugh2,McGaugh5,Dutton} and velocity dispersion profiles of elliptical galaxies \citep{Chae}. However, these benchmark values are ruled out at $5\sigma$ based on the MWRC slope with the fiducial values. Besides, even if we consider the concentrated baryonic disk model, the values for $p \ge 1$ and $\delta \ge 1$ are still ruled out at more than $5\sigma$ (see Fig.~2).

Apart from using the fiducial values of the baryonic parameters, we consider conservative ranges of $M_d$ and $a_d$ for a more rigorous investigation. Note that the room for varying the value of $M_b$ is very small because it is mainly determined by the very inner RC data, which is almost independent of MOND or any particular dark matter model. Therefore, we will still adopt $M_b=1.55 \times 10^{10}M_{\odot}$. Besides, previous studies have shown that MOND favors small values of $a_d$ \citep{Gerhard,McGaugh5}. The RC data suggest $a_d \sim 2-4$ kpc and $M_d \sim (2.9-5.4) \times 10^{10}M_{\odot}$ under the MOND framework \citep{McGaugh5}. These ranges are generally consistent with the mass model predictions in many studies \citep{Sofue,McGaugh4,Ou}. To perform a conservative test, we allow wider ranges of $a_d$ and $M_d$: $a_d=1-10$ kpc and $M_d=(1-10)\times 10^{10}M_{\odot}$.

In Fig.~3, we plot the RC gradient $dV/dR$ as a function of $a_d$ for different $M_d$. Since most of the previous studies assume $p \ge 1$ and $\delta \ge 1$ \citep{Dutton,McGaugh5}, we only consider the simplest form (the benchmark values) of $p=1$ and $\delta=1$ for both interpolating families because larger values of $p$ and $\delta$ would make the RC gradient larger. We find that almost all RC gradients are larger than the $5\sigma$ upper bound of the observed RC gradients, except for the $p$-family with the extreme parameter $M_d=10^{11}M_{\odot}$. This indicates that a large parameter space of $a_d$ and $M_d$ in the conservative ranges are ruled out at $5\sigma$ for $p \ge 1$ and $\delta \ge 1$.

\section{Discussion}
In this article, we present the RC gradient analysis for MOND using the MWRC data at large $R$. Thanks to the high-quality measurements, the MWRC data have very small error bars at large $R$, which give an excellent platform to test modified gravity theories or constrain dark matter model. The MWRC data show a very clear declining trend at large $R$ which potentially challenges MOND because it predicts an almost flat RC at large $R$. The $5\sigma$ range of the RC gradient is $dV/dR=-5.07^{+2.51}_{-2.49}$ km/s/kpc (without considering the systematic uncertainties). On the other hand, by considering two major families of the interpolating functions, we have derived the RC gradient formulas explicitly for the MOND framework. We first follow the benchmark baryonic disk density profile and use the fiducial values of $a_d$, $M_d$ and $M_b$ to calculate the RC gradient. There exist $5\sigma$ allowed ranges of $p=0.19-0.46$ and $\delta=0.17-0.43$. However, when these are combined with the Mercury precession constraints $p \ge 1.22$ and $\delta \ge 0.33$ \citep{Chan2}, only a narrow range of $\delta$ remains. This suggests that almost all possible variations of MOND are ruled out. Even in the extreme case (concentrated baryonic disk model), the benchmark MOND parameters $p \ge 1$ and $\delta \ge 1$ are still ruled out beyond $5\sigma$.

Besides, we also consider possible conservative ranges of $M_d=(1-10) \times 10^{10}M_{\odot}$ and $a_d=1-10$ kpc. For the benchmark cases $p=1$ and $\delta=1$, almost all of the RC gradients calculated are above the $5\sigma$ upper bound of the observed RC gradient. As a larger $p$ or $\delta$ gives less negative RC gradients, this gives a severe challenge to the MOND phenomenology in our Galaxy. Nevertheless, we did not consider the systematic uncertainties of the RC data points shown in \citet{Ou}. Some of the systematic uncertain factors (e.g. the solar position) would somewhat systematically underestimate or overestimate the values of RC, without affecting the value of the RC gradient. However, it is not sure how the other factors discussed in \citet{Ou} would affect the RC gradient calculation. If we assume 1\% systematic uncertainties and combine with the observational uncertainties, the $5\sigma$ range of the RC gradient would enlarge to $dV/dR=-5.07^{+4.14}_{-3.77}$ km/s/kpc. However, if we focus on the $2\sigma$ range instead, similar parameter space of the MOND parameters are still ruled out at $2\sigma$ (95.45\% C.L.) (see Fig.~3).

This is the first time we can rule out a large parameter space of galactic MOND phenomenology at more than $5\sigma$. In fact, many previous studies, like the SPARC analysis, have claimed that MOND works very well in galaxies \citep{McGaugh,Islam}. The radial acceleration relation shown in SPARC gives an excellent agreement with MOND's prediction \citep{McGaugh,McGaugh2}. However, the rotation curve data for the galaxies outside our Local group have large uncertainties. Moreover, the fittings also involve some unknown parameters like the mass-to-light ratio. Therefore, there is plenty of room for MOND to fit with the data. In fact, some later studies using better analysis tools have challenged the existence of a universal acceleration scale $a_0$ \citep{Rodrigues,Marra,Chan3} and the core-cusp nature in galaxies \citep{Eriksen} predicted by the MOND phenomenology. 

Fortunately, the very small error bars of the MWRC generated from the analysis combining the data of APOGEE DR17, GAIA, 2MASS and WISE can give a robust test for the galactic MOND phenomenology. We have tested two major families of interpolating functions which can almost represent all possible transition functions between the deep-MOND and Newtonian regimes. We have also considered wide conservative ranges of $a_d$ and $M_d$. Our robust analysis gives a severe challenge to the MOND phenomenology in our Galaxy, unless the effect of the systematic uncertainties is larger than our expectation. The phenomenological MOND might be just a rough approximation and it is definitely not universal for all galaxies. Besides, the small uncertainties of the MWRC generated can also be used to analyze the radial acceleration relation for our Galaxy. This can further examine the MOND phenomenology and the alleged universal acceleration scale shown in galaxies.

\begin{figure}
\vskip 10mm
 \includegraphics[width=130mm]{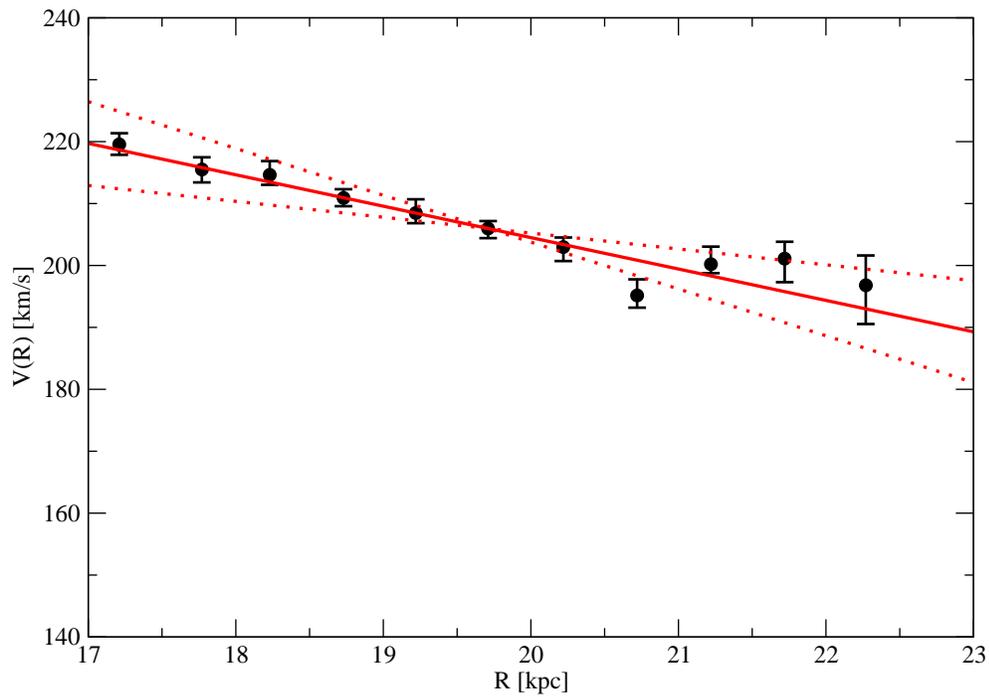}
 \caption{The black dots with error bars represent the data of the MWRC for $R=17-23$ kpc \citep{Ou}. The red solid line indicates the best-fit RC gradient $dV/dR=-5.07$ km/s/kpc. The red dotted lines indicate the $5\sigma$ upper and lower bounds of the best-fit RC gradients.}
\vskip 10mm
\end{figure}

\begin{figure}
\vskip 10mm
 \includegraphics[width=140mm]{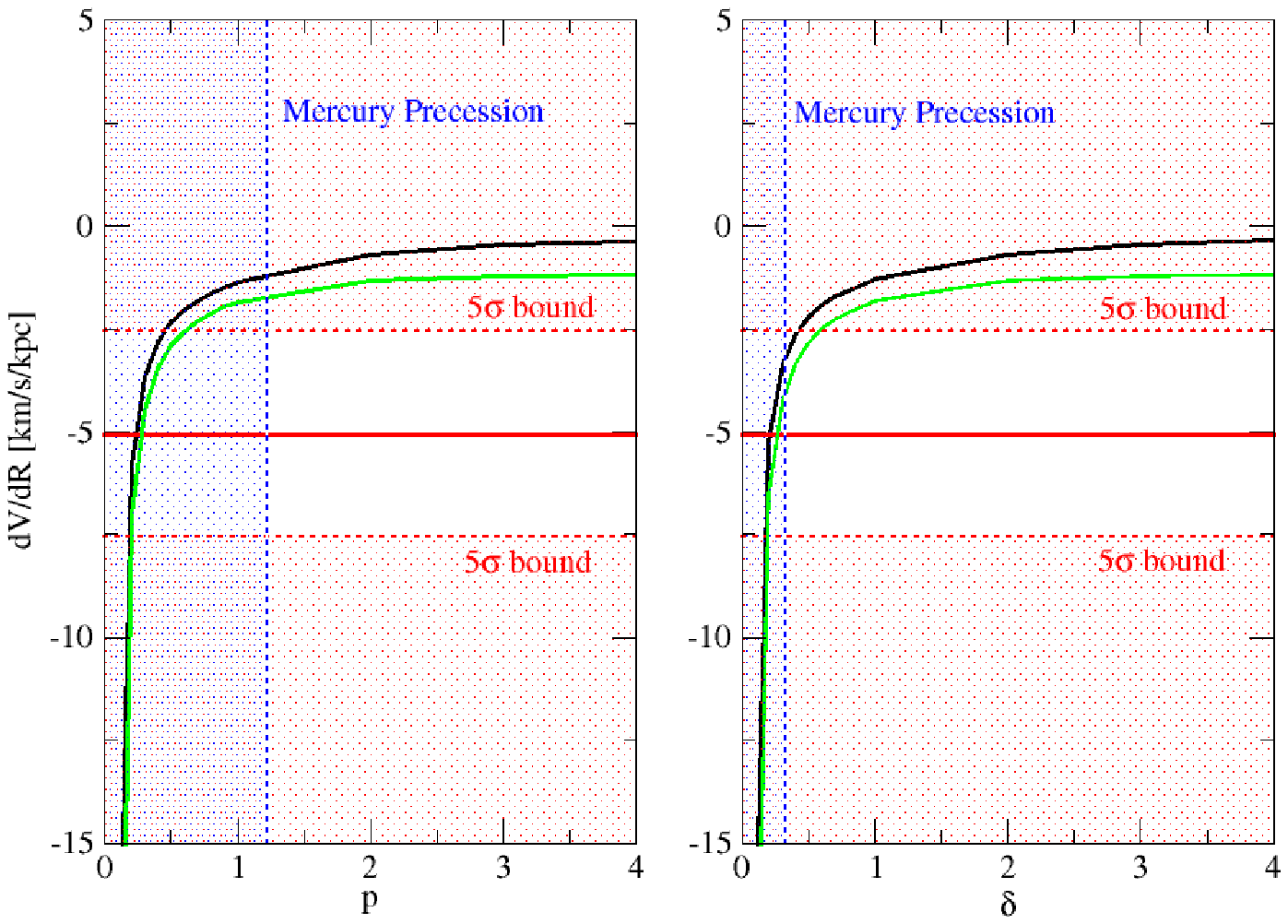}
 \caption{The black lines (assuming an exponential baryonic disk) and green lines (assuming a small baryonic disk with radius $<17$ kpc) indicate the functions of $dV/dR$ against $p$ (left) and $\delta$ (right). The red solid line indicates the best-fit RC gradient and the shaded regions bounded by the red dotted lines represent the parameter space ruled out at $5\sigma$ based on the MWRC data. The blue shaded regions indicate the parameter space ruled out at $1\sigma$ based on the Mercury precession data \citep{Chan2}.}
\vskip 10mm
\end{figure}

\begin{figure}
\vskip 10mm
 \includegraphics[width=140mm]{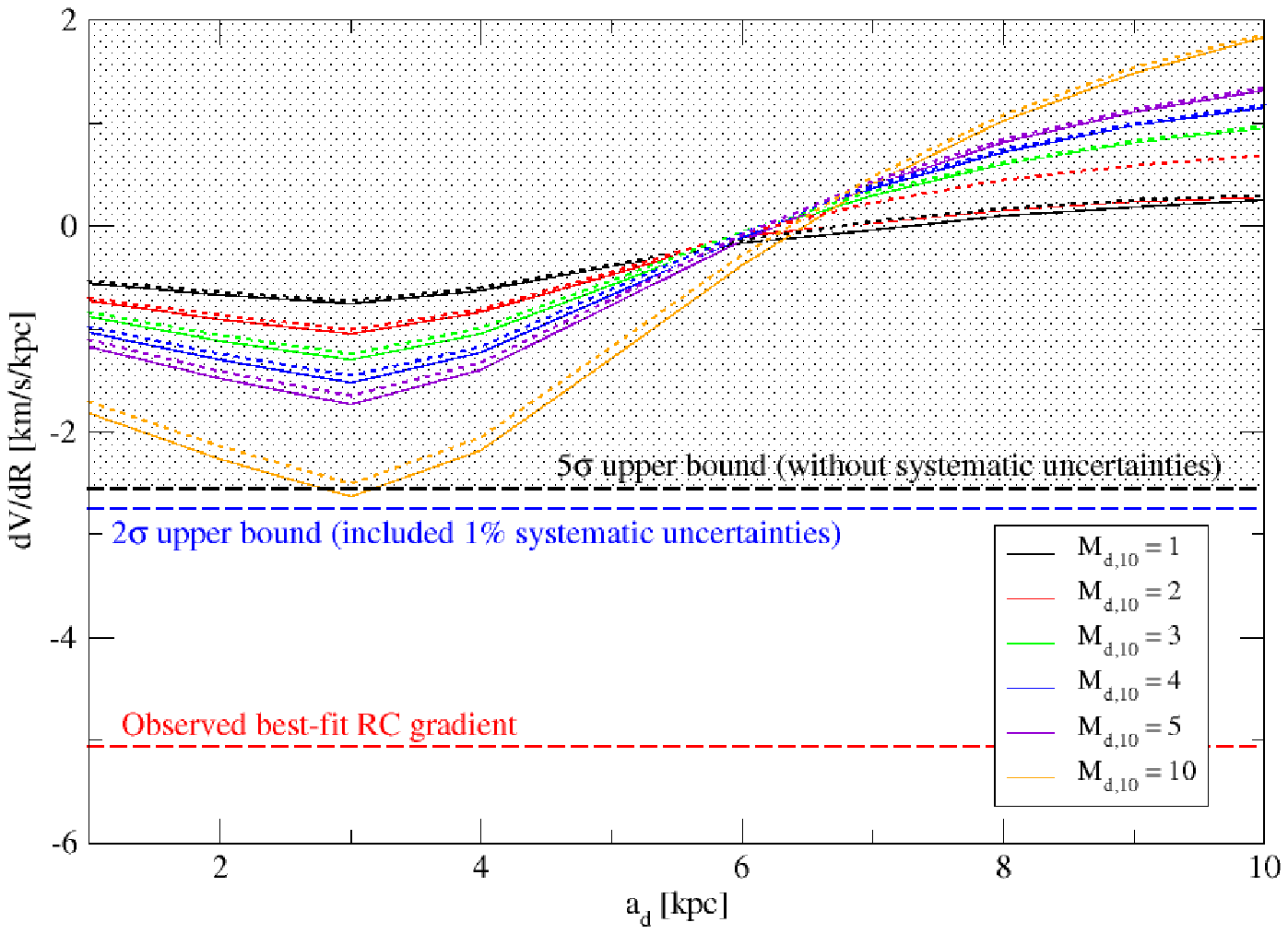}
 \caption{The colored lines (solid: $p$-family; dotted: $\delta$-family) represent the MOND RC gradient $dV/dR$ against the disk scale length $a_d$ for different disk total mass $M_{d,10}$, where $M_{d,10}=M_d/10^{10}M_{\odot}$. The red, black, and blue dashed lines respectively indicate the best-fit RC gradient, its $5\sigma$ upper bound (without considering systematic uncertainties), and its $2\sigma$ upper bound (including 1\% systematic uncertainties) from the MWRC data. The shaded region is the region ruled out by the $5\sigma$ upper bound.}
\vskip 10mm
\end{figure}

\section{Acknowledgements}
The work described in this paper was partially supported by a grant from the Research Grants Council of the Hong Kong Special Administrative Region, China (Project No. EdUHK 18300922).

\end{document}